**Understanding Immune Dynamics in Liver Transplant Through Mathematical Modeling**

Julia Bruner[1]*, Kyle Adams[2], Skylar Grey[3], Mahya Aghaee[3], Sergio Duarte[1], Ali Zarrinpar[1], Helen Moore[3]*

*Corresponding Authors, juliabruner@ufl.edu, helen.moore@medicine.ufl.edu

1. Department of Surgery, College of Medicine, University of Florida, Gainesville, Florida, USA.
2. Department of Mathematics, College of Liberal Arts and Sciences, University of Florida, Gainesville, Florida, USA.
3. Department of Medicine, College of Medicine, University of Florida, Gainesville, Florida, USA.

**ORCIDs:** JB 0000-0002-6790-5096, KA 0009-0006-2846-1564, SG 0000-0003-0852-0070, SD 0000-0003-1036-6850, AZ 0000-0002-6812-1816, HM 0000-0002-2185-2785

## Abstract

Liver transplant can be a life-saving procedure for patients with end-stage liver disease. With the introduction of modern immunosuppressive therapies, short-term survival has significantly improved. However, long-term survival has not substantially improved in decades. Consequently, causes of death are now more likely to be due to the toxicities and side-effects of long-term immunosuppression rather than rejection. In order to study the balance of immunosuppression and rejection, we developed the first mechanistic mathematical model of liver transplant and immune system dynamics. We determined key cells and interactions in the model using literature information; we then used sensitivity analysis to determine key pathways driving the health status of the transplanted liver. We found that dynamics related to cytotoxic T cells and IL-2, in addition to the liver itself, are key determinants of liver graft injury. This has significant implications for the use of tests to monitor patients, and therapeutic strategies to prevent or treat liver transplantation rejection. Future work to collect appropriate data and parametrize the model would be valuable in improving our understanding of the dynamics of this system. We also note that our model could be tailored to model transplant of other organs.

## Introduction

Liver transplantation is the gold standard treatment for end-stage liver disease. In 2022, 37,436 liver transplants were performed worldwide, an 8% increase from 2021.[1] In 2023, 10,660 liver transplants were performed in the US alone, a record-breaking number marking growth for the 11th straight year.[2] However, as the number of liver transplants continues to rise, a lack of improvement in long-term outcomes remains a significant limitation. While early outcomes after liver transplantation have markedly improved over the past 3 decades, this trend has not been matched by long-term outcomes.[2] In 2022, the average 10-year survival among liver transplant patients was less than 65%.[3] The lack of improvement in mortality reflects the ongoing challenge of immunosuppression management, evidenced by a changing landscape in causes of mortality. Among the most frequent causes of death are infection (5.78%), *de novo* malignancy (3.48%), cardiovascular disease (2.71%) while graft rejection represents just 0.36% of 10-year mortality.[4] Hence strides made to reduce rejection-related graft loss and mortality have been offset by the complications of aggressive immunosuppression. Given that both under- and over-immunosuppression carry substantial consequences for long-term outcomes, better understanding and better strategies are needed to help clinicians successfully balance these risks throughout long-term dosing.

Mechanistic mathematical models have been used previously to help understand disease settings where the balance between states is important, and to make decisions based on this understanding. For example, such models have been used extensively to study glucose-insulin dynamics,[5–12] as well as coagulation cascade dynamics.[13–18] These models help by combining and synthesizing all known relevant information in a quantitative model of the dynamics, informing and improving dosing decisions. They enable the application of powerful analytical methods to explore the models and understand more about the model behavior. And they can be used for simulating virtual clinical trials, testing different regimens, and asking "what if" questions before experimental or clinical studies are conducted.[19,20] Furthermore, they can be used to optimize drug regimens.[21]

More specifically, a variety of mathematical models have been proposed to study transplant immunology. Stegall and Borrows reviewed methods to predict graft survival in a renal transplant setting, and strongly advocated for mechanism-based modeling studies of graft rejection.[22] Raimondi et. al discussed a wide variety of models, including statistical models, mechanistic models, and bioinformatics approaches to studying transplant immunology, and encouraged additional collaborations between experimentalists and math modelers.[23] Fribourg reviewed advantages of applying previously-published models to transplant settings to understand immune responses, illustrating this with an example.[24]
Markovska developed and analyzed a system of ordinary differential equations (ODEs) to model the adaptive immune response after organ transplantation.[25] An used an agent-based model to explore immune dynamics in a general solid organ transplant setting.[26] Best et. al explored immune tolerance by studying interactions between T cells and APCs through an optimization-based model, with the goal of understanding transplant organ tolerance.[27] Day et. al used an ODE model to study immune tolerance in a solid organ transplant setting, specifically focusing on ischemia/reperfusion injury.[28]

Ciupe et al. established one of the earliest ODE models that investigates immune dynamics in a transplant setting.[29] They fit a stochastic model to data to better understand dynamics about T cells in a thymus transplant, including information about TCR-specific and TCR-nonspecific regulatory signals. Gateno et al. developed an ODE model to describe immune system dynamics in a general setting of any rejection of a solid organ transplant.[30] In their model, they incorporate a treatment, cyclosporine, to investigate cellular responses of the immune system as well as the effects of various doses of treatment. Arciero et al. established a model in a murine heart transplant rejection setting with ODEs[31] and then Lapp et al. used it to specifically focus on how adoptive transfer of Tregs can affect graft survival.[32] This model, which used experimental data to calibrate T cell populations, also includes a dosing function to consider different doses of Tregs. Banks et al. established a model to describe the immune response of renal transplant recipients, and applied optimal control to determine optimal doses of immunosuppression and antiviral drugs.[33] Many of these authors call for additional mechanistic mathematical modeling of organ transplant and rejection settings to help improve patient outcomes. We agree with this sentiment,[34] and in this work, we contribute with our own mechanistic mathematical model of immune dynamics in a liver transplant setting. Unique aspects of our model include rigorous parameter derivations from the literature, the specific cell and cytokine populations chosen, including IL-2, and our model's specificity to the liver, as opposed to a general solid organ.

In this work, we created a novel mathematical model describing mechanisms involved in acute rejection taking place approximately one year after liver transplantation. We used an extensive literature search and our knowledge of cellular interactions to derive information at the organ rejection level. We applied sensitivity analysis to determine the most important factors that drive rejection. Our ultimate aim is to better understand transplant-immune system dynamics at the cellular level, especially the key cells and interactions involved. In future work, we plan to fit this model to data and validate it. We could then use the model to mathematically optimize immunosuppression dosing, determining doses high enough to prevent rejection, without being excessively high. This approach could help prevent both rejection and over-immunosuppression related complications, which continue to limit long-term outcomes.

## The Model

### Overview

We constructed our model of rejection based on interactions between the transplanted liver (i.e., allograft), antigen-presenting cells (APCs), helper T cells ($T_H$), cytotoxic T cells ($T_C$), regulatory T cells ($T_R$), and interleukin-2 (IL-2). Importantly, our model details the interaction of circulating immune cells with the allograft. These components are believed to be responsible for some of the most-significant dynamics driving the rejection process in the long-term setting. In the next section, we provide more details about each of these and their role in the dynamics. The relevant interaction pathways, labeled in the model diagram in **Fig. 1**, are indicated in parentheses. These interactions are summarized in **Table 1**.

### Model Components and Justification for Their Inclusion
<u>Liver Allograft ($L$)</u>. **The liver allograft plays an active role in the immunological process of rejection, through its release of antigens.** After transplantation, major histocompatibility complex (MHC) molecules, as well as minor-histocompatibility antigens, are shed from the donor organ cells into the organ recipient's blood (pathway **b**). This is evidenced by the appearance of allograft-derived soluble human leukocyte antigen (sHLA) class I antigens in recipient blood circulation shortly after transplantation, which can continue for years.[35–38] High levels of graft-derived class I sHLA are correlated with acute rejection, as well as other complications.[35–38] While donor-recipient HLA matching can be used to mitigate this process, resulting in improved graft survival and reduced acute rejection in other organs,[39] this has not been found to be helpful in liver transplantation. Hence, both major and minor histocompatibility antigens contribute to the rejection immune response. Major and minor histocompatibility antigens may be either directly recognized by T cells or phagocytosed, processed, and presented to T cells by antigen-presenting cells (pathway **g**), activating the recipient immune response (pathways **m**).[40–42] Given high regenerative capacity,[43] in a stable state, the liver is presumed to be in homeostasis where the rate of replication of mature hepatocytes (represented by the brown circular arrow at the top of *Liver* in **Fig. 1**) is approximately equal to the natural death rate (pathway **h**). Upon death, whether natural loss or immune-mediated loss during rejection, allograft hepatocytes release enzymes aspartate transaminase (AST) and alanine transaminase (ALT), and the allograft cholangiocytes release alkaline phosphatase (ALP) into circulation (pathways **c**,**d**,**f**).[44] We assume that each of these is proportional to the loss rate for these cells from the liver, which account for the majority of cells in the liver. The circulating levels of AST, ALT, and ALP are easily measured, which can be used in the future to tie our model to data. However, our real interest is in the level of healthy liver allograft hepatocytes. Thus we will use the level of healthy hepatocytes at day 30, $L(30)$, as our quantity of interest (QOI) in later calculations.

**Antigen-Presenting Cells (*A*). APCs activate adaptive cellular immune responses after transplantation through three distinct mechanisms - direct, indirect, and semi-direct allorecognition (recognition of "other").** In direct allorecognition, donor APCs present their allogeneic MHC directly to recipient T cells. In indirect allorecognition, donor MHC and other antigens released from the graft are phagocytosed, processed, and presented by recipient APCs to the recipient T cells. Finally, in semi-direct allorecognition, recipient APCs obtain and present intact donor MHC on their surface to recipient T cells.[45–47] The direct allorecognition pathway is generally believed to drive early acute rejection events, while indirect allorecognition is thought to drive late rejection events and chronic rejection.[48,49] However the relative contributions of these three mechanisms to rejection remains unclear. Hence, in our initial model, the antigen presentation encompasses all three allorecognition pathways with APCs representing a general population of both donor and recipient cells (pathway **a**) that are loaded with antigen shed by the donor organ. Pathway **b**, mentioned above, is required for the formation/recruitment of these loaded APCs, which we represent by showing pathway **a** with a dashed circle, indicating dependence on pathway **b**. The natural loss of APCs, i.e., as they reach the end of a finite lifespan, is represented by pathway **i**.

**Helper T cells ($T_H$). Alloreactive CD4+ helper T cells become activated upon recognition of donor antigens processed and presented on class II MHCs by APCs.** Once activated by APCs (pathway **m**, which requires interaction with APCs, represented by pathway **g**), CD4+ T cells play a role in organ rejection through a variety of mechanisms that support CD8+ T cell differentiation, including proinflammatory cytokine secretion, direct cell-to-cell contact, and upregulation of costimulatory signals on APCs (pathway **l**).[50–52] Evidence for differentiation to defined helper cell lineages Th1, Th2, and Th17 is variable, with all three subtypes reportedly contributing to rejection mechanisms depending on the setting.[53,54] Regardless of lineage, interaction of CD4+ with CD8+ T cells is required for maximal rejection (pathway **k**).[55,56] Helper T cells have a finite lifespan. Natural loss at completion of this lifespan is represented by pathway **y**.

**Cytotoxic T cells ($T_C$). Activated CD8+ cytotoxic T cells are the central mediators of the cellular immune response in acute rejection.** Recognition of alloantigens in the presence of costimulatory signals and cytokines from APCs and CD4+ helper T cells generates armed effector cytotoxic T cells which migrate to the allograft and induce the death of donor cells. (pathway **e**, shown as increasing the loss rate **h**)[57,58] In this setting, CD8+ T cells induce apoptosis primarily through the granzyme/perforin pathway.[59] Cytotoxic granules containing the membrane-disrupting protein perforin and serine proteases called granzymes are secreted, triggering targeted cell death.[59] As a side note, quantified perforin and granzyme B (GrB) mRNA levels have been used experimentally as non-invasive biomarkers to monitor acute rejection in kidney transplantation, though the challenges in quantifying these levels has prevented these from being used clinically.[60] The role of cytotoxic T cells in rejection is also observed clinically in associations between elevated CD8+ T cells in subset peripheral blood counts and increased frequency of allograft dysfunction.[61,62] Furthermore, mouse models of minor histocompatibility mismatch demonstrate that depletion of CD8+ T cells prevents allograft rejection, though this result varies with the transplanted tissue.[63,64] Cytotoxic T cells have a finite lifespan. Natural loss at completion of this lifespan is represented by pathway **t**.

**Regulatory T cells ($T_R$). Regulatory T cells, (Tregs), oppose the proinflammatory responses generated by APC, $T_C$, and $T_H$ interactions to promote immune tolerance.** While Tregs are generally divided into thymically-derived and peripherally- induced populations, it is unclear which of these subsets is most responsible for suppression of the allogeneic immune response after transplantation (pathway **v**).[65] In this initial model, we represent Tregs as a single population of CD4+CDC25+FoxP3+ cells specific for the allograft organ. Tregs may exert suppressive function through constitutive expression of CD25, the high-affinity IL-2 receptor, which sequesters IL-2 (pathway **u**) and other common gamma chain-binding proinflammatory cytokines from conventional (i.e., cytotoxic and helper) effector T cells.[66,67] Tregs also achieve tolerogenic effects via trogocytosis of APC-expressed costimulatory molecules such as CD80/CD86 (B7), a process in which Tregs physically deplete these molecules from APCs during direct cell-to-cell contact (pathway **j**).[66,68] Alternative mechanisms of suppression include secretion of anti-inflammatory cytokines, hydrolysis and depletion of ATP, and granzyme-dependent pathways, among others.[69–71] Given this wide variety of suppressive functions, high ratios of Tregs to conventional T cells are preventative against allograft rejection and promote tolerance.[72,73]

**Interleukin-2 ($I$). IL-2 is an immunomodulatory cytokine critical to both inflammatory and tolerogenic immune responses following transplantation.** Activated helper and cytotoxic T cells produce IL-2 (pathways **n** and **o**), which acts as a potent autocrine and paracrine signal to stimulate T cell proliferation (pathways **p** and **s**).[74,75] Additionally, IL-2 stimulates expression of perforin and granzyme by cytotoxic T cells, and supports differentiation of CD4+ helper T cells to proinflammatory lineages.[74] IL-2-driven proliferation of conventional T cells requires

internalization of IL-2 (pathways **q** and **r**, required for proliferation pathways **p** and **s**)[75,76]. Inhibition of internalization, without disruption of IL-2 binding to the IL-2 receptor complex, is sufficient to block T cell proliferation.[75] However, in vivo, depletion of IL-2 or IL-2 receptors results in autoimmune phenotypes, highlighting the equal importance of IL-2 to the promotion of tolerance.[77,78] Unlike conventional T cells, Tregs do not produce IL-2. However, Tregs require continuous IL-2 signaling for survival (pathway **u**), maintenance of FoxP3 expression, and to support suppressive functions (pathway **x**).[79–81] This is evidenced by reduced in vitro suppressive capacity of human Tregs upon deletion of the IL-2 receptor alpha gene using CRISPR/Cas9 gene editing.[82] Just as for the stimulation of proliferation in conventional T cells, IL-2-dependent effects in Tregs occur through signaling processes involving internalization (pathway **u**). In vitro and in vivo, mutant IL-2 receptor complexes that enhance Treg IL-2 internalization and recycling, prolong Treg suppressive functions and promote proliferation.[83,84] Natural degradation of IL-2 is represented by pathway **w.** Given both pro- and anti-inflammatory capacities, IL-2 represents a critical point of feedback between activation and tolerance, though the exact balance of these actions remains to be determined. In mouse models, low-dose IL-2, which favors tolerogenic effects on Tregs has been shown to prevent chronic cardiac allograft rejection and prolong allograft survival.[85,86] In contrast, in a pilot clinical trial of stable liver transplant recipients, low dose IL-2 failed to promote tolerance or allow for the discontinuation of immunosuppressive medications. Instead, IL-2 induced an inflammatory response resembling T cell-mediated rejection (TCMR) despite an increase in circulating Tregs. In this model, we consider both pro- and anti-inflammatory effects of IL-2.[87]

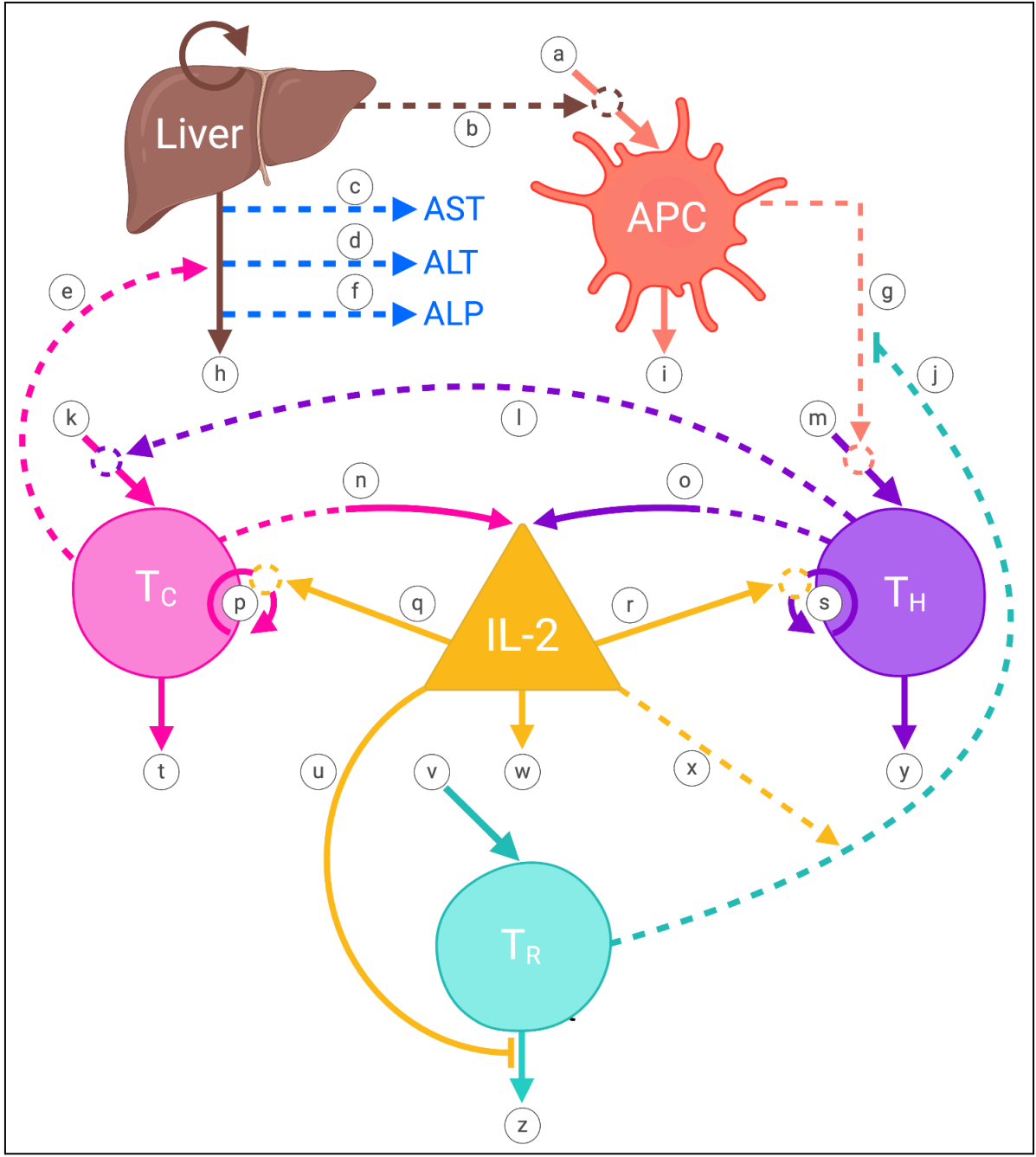

**Figure 1: Model Diagram - Dynamics of Acute Rejection in Liver Transplantation.**
Visualization of the mathematical model. The liver allograft, each cell population, and the cytokine interleukin-2 are each represented by labeled shapes with the following respective abbreviations: liver allograft (L), antigen-presenting cells (APC), helper T cells ($T_H$), cytotoxic T cells ($T_C$), and regulatory T cells ($T_R$), and interleukin-2 (IL-2), respectively. Each labeled arrow (**a**-**z**, labeled from top to bottom, left to right) represents a pathway described in the *Model Pathways* section. Dashed arrows represent enhancing interactions between model components, with the dashed arrow leaving the population responsible for the enhancement and the arrowhead pointing to the affected pathway. Lines ending in a perpendicular line segment represent inhibitory actions with the line leaving the population responsible for the inhibition, and the line segment parallel to the affected pathway (pathways **j** and **u**). Solid arrows (or inhibitory line segments, i.e., pathway **u**) indicate a change in population number with inward-pointing arrows representing an increase in number and outward-pointing arrows representing a decrease in number. More specifically, solid arrows pointing straight down indicate loss or degradation, circular solid arrows represent proliferation, and diagonal arrows pointing inward to model components represent a source. Solid source arrows interrupted by dashed circles (pathways **a**, **k**, **m**, **p**, and **s**) represent the requirement of the dashed pathway interactions for occurrence of the solid line pathway. Please see *Model Pathways* for additional information on required interactions. Pathways that are dashed leaving a model component agent but which become solid entering another model component (pathways **n** and **o**) indicate production of the affected model agent, IL-2, without reduction in the responsible cell population, $T_C$ and $T_H$ respectively. Proliferation of allograft hepatocytes (top left circular arrow) is unlabeled as this dynamic is not represented mathematically in the model.

| Table 1: Summary of Model Pathways ||
|---|---|
| **Label** | **Description** |
| **a** | Source of alloantigen-presenting APCs |
| **b** | Alloantigens are shed from liver allograft |
| **c** | AST is released upon damage to liver allograft hepatocytes |
| **d** | ALT is released upon damage to liver allograft hepatocytes |
| **e** | Activated cytotoxic T cells attack liver allograft |
| **f** | ALP is released upon damage to liver allograft hepatocytes |
| **g** | Alloantigen-presenting APCs activate alloreactive helper T cells |
| **h** | Natural loss of allograft hepatocytes |
| **i** | Natural loss of APCs |
| **j** | Tregs suppress activation of alloreactive helper T cells by APCs |
| **k** | Source of activated cytotoxic T cells |
| **l** | Helper T cells participate in activation of cytotoxic T cells |
| **m** | Source of activated helper T cells |
| **n** | Production of IL-2 by activated cytotoxic T cells |

| | | |
|---|---|---|
| o | Production of IL-2 by activated helper T cells | |
| p | Proliferation of cytotoxic T cells | |
| q | IL-2 leads to proliferation of cytotoxic T cells | |
| r | IL-2 leads to proliferation of helper T cells | |
| s | Proliferation of helper T cells | |
| t | Natural loss of cytotoxic T cells | |
| u | Tregs remove IL-2 from environment, promoting Treg survival | |
| v | Source of regulatory T cells | |
| w | Natural turnover of IL-2 | |
| x | IL-2 boosts Treg suppression of APC activation of helper T cells | |
| y | Natural loss of helper T cells | |
| z | Natural loss of regulatory T cells | |

| Table 2: Variable Definitions and Initial Values | | |
|---|---|---|
| Variable | Definition | Initial Value |
| L | Healthy liver hepatocytes | 2.0 x $10^{11}$ cells |
| A | Antigen presenting cells (APCs) | 8.0 cells/µL |
| $T_H$ | Activated helper T cells | 70 cells/µL |
| $T_C$ | Activated cytotoxic T cells | 100 cells/µL |
| $T_R$ | Regulatory T cells (Tregs) | 1.31 cells/µL |
| I | Interleukin 2 (IL-2) | 0.0113 ng/µL |

**Model Equations**

(1) $\dfrac{dL}{dt} = - \overbrace{\delta_L L}^{h} \Big( \overbrace{\dfrac{\alpha_{CL} T_C}{\beta_{CL} + T_C}}^{e} \Big)$

$$(2) \quad \frac{dA}{dt} = \lambda_L \overbrace{\delta_L L}^{h} \overbrace{(1 + \frac{\alpha_{CL} T_C}{\beta_{CL} + T_C})}^{e} - \overbrace{\delta_A A}^{i}$$

$$(3) \quad \frac{dT_H}{dt} = (\overbrace{\frac{\alpha_{AH} A}{\beta_{AH} + A}}^{g,m}) \left[ 1 - (\overbrace{\frac{\alpha_{RA} T_R}{\beta_{RA} + T_R}}^{j})(\overbrace{1 + \frac{\alpha_{IRA} I}{\beta_{IRA} + I}}^{x}) \right] + \overbrace{\gamma_H T_H (1 - \frac{T_H}{K_H})}^{s} \overbrace{(\frac{\alpha_{IH} I}{\beta_{IH} + I})}^{r} - \overbrace{\delta_H T_H}^{y}$$

$$(4) \quad \frac{dT_C}{dt} = \overbrace{\frac{\alpha_{HC} T_H}{\beta_{HC} + T_H}}^{k,l} + \overbrace{\gamma_C T_C (1 - \frac{T_C}{K_C})}^{p} \overbrace{(\frac{\alpha_{IC} I}{\beta_{IC} + I})}^{q} - \overbrace{\delta_C T_C}^{t}$$

$$(5) \quad \frac{dT_R}{dt} = \overbrace{s_R}^{v} - \overbrace{\delta_R T_R}^{z} \overbrace{(1 - \frac{\alpha_{IR} I}{\beta_{IR} + I})}^{u}$$

$$(6) \quad \frac{dI}{dt} = \overbrace{\frac{\alpha_{CI} T_C}{\beta_{CI} + T_C}}^{n} + \overbrace{\frac{\alpha_{HI} T_H}{\beta_{HI} + T_H}}^{o} - \lambda_C \overbrace{\gamma_C T_C (1 - \frac{T_C}{K_C})}^{p} \overbrace{(\frac{\alpha_{IC} I}{\beta_{IC} + I})}^{q} - \lambda_H \overbrace{\gamma_H T_H (1 - \frac{T_H}{K_H})}^{s} \overbrace{(\frac{\alpha_{IH} I}{\beta_{IH} + I})}^{r} - \lambda_R \overbrace{\delta_R T_R}^{z} \overbrace{(1 - \frac{\alpha_{IR} I}{\beta_{IR} + I})}^{u} - \overbrace{\delta_I I}^{w}$$

**Parameter Constraints**

We note all parameters are assumed to be positive. We also assume that $\alpha_{IR} \leq 1$; otherwise, the **u** pathway could contribute a negative factor, which would result in the natural loss of IL-2 becoming a gain of IL-2 in Equation (5). We also assume that $\alpha_{RA}(1 + \alpha_{IRA}) \leq 1$; this ensures that all quantities have the appropriate sign in Equation (3).

**Equation Descriptions**

**Equation (1)** models the rate of change of hepatocytes from the liver with respect to time. We assume that the liver is initially in homeostasis. Mathematically, this means that the production and death rates of liver hepatocytes are equal, so that if $\frac{dL}{dt} = s_L L - \delta_L L$, then $s_L = \delta_L$. The effect of cytotoxic T cells on the loss of liver hepatocytes is represented by pathway **e**. We chose to represent this loss with a Michaelis-Menten term, so that the rate is limited. This yields $\frac{dL}{dt} = s_L L - \delta_L L(1 + \frac{\alpha_{CL} T_C}{\beta_{CL} + T_C})$. Using the relationship $s_L = \delta_L$, this simplifies to $\frac{dL}{dt} = -\delta_L L(\frac{\alpha_{CL} T_C}{\beta_{CL} + T_C})$.

**Equation (2)** models the rate of change of recipient alloantigen-presenting APCs with respect to time. The main source of these APCs is the alloantigen released by the loss of hepatocytes, labeled as pathway **h**. This is boosted by cytotoxic T cells attacking hepatocytes, labeled as pathway **e**. This source term is multiplied by $\lambda_L$, a constant of proportionality that takes into account multiple factors described below. We assume there is an excess of APCs available to be loaded with antigen, and so this source just depends on the loss rate of hepatocytes, which is itself proportional to the number of hepatocytes. The source of alloreactive APCs is dependent on alloantigen release from hepatocyte loss, as alloantigen is necessary to facilitate phagocytosis and presentation. These processes transform resting APCs into alloreactive APCs. Since liver donors typically do not have high homology (matched HLA) between recipient and donor, we assume that, despite high volume of alloantigen release, this process is antigen-limited. We use a Michaelis-Menten style term for this dependency. The loss term, labeled as pathway **i**, represents the natural loss of APCs, which we assume is proportional to the concentration of APCs present.

**Equation (3)** models the rate of change of activated helper T cells with respect to time. Without APCs, alloreactive helper T cells will not be activated, and this is reflected by the term labeled **g,m**. This process is limited by Tregs suppressing APC presentation, which is reflected in pathway **j**. Pathway **j** is multiplied by pathway **x**, which represents the boost from IL-2 in Treg suppression of APC presentation. The fractions within terms labeled **g,m, j,** and **x** are modeled after Michaelis-Menten equations. Pathway **s** represents the logistic growth of helper T cells from natural proliferation, which is multiplied by pathway **r**, which represents the necessary effect IL-2 has on T cell proliferation. We note that the growth of helper T cells from IL-2 is proportional to the loss of IL-2 from helper T cells in **Eq. (3)**. The natural loss of helper T cells, pathway **y**, is proportional to the population of helper T cells.

**Equation (4)** models the rate of change of activated cytotoxic T cells with respect to time. The first term, labeled **k,l**, models the activation of naive CD8+ T cells by interaction with antigen and activated helper T cells. We use a Michaelis-Menten term to ensure the rate stays bounded. Pathway **p** represents the natural logistic proliferation of cytotoxic T cells, which is multiplied by pathway **q**, which represents the requirement of IL-2 for cytotoxic T cell proliferation. We note this term is proportional to the **p*q** loss term in equation 6. Term **t** represents the natural loss of cytotoxic T cells and is proportional to the cytotoxic T cell population.

**Equation (5)** models the rate of change of Tregs with respect to time. We assume a constant source rate of Tregs being recruited from outside of our model**.** The factor **z** represents the natural loss of Tregs. The factor **u**, which is less than or equal to one, represents the lifespan-lengthening effect IL-2 has on Tregs. We use a Michaelis-Menten term to ensure a bound on this effect size.

**Equation (6)** models the rate of change of IL-2 with respect to time. There are two source terms: pathway **n** represents the production of IL-2 from cytotoxic T cells and pathway **o** represents the production of IL-2 from helper T cells. These terms have limits, even if the cell populations producing IL-2 get very large. We assume that other production of IL-2 is negligible in comparison. The loss terms with factors of $\lambda_C$, $\lambda_H$, and $\lambda_R$ represent the decrease in IL-2 levels due to binding to and internalization by cytotoxic, helper, and regulatory T cells, respectively. We will describe the IL-2 loss term related to $T_C$, but the loss terms due to $T_H$ and $T_R$ are similar. $T_C$ internalization and consumption of IL-2 leads to new $T_C$ production, which appears in **Eq. (4)** as **p*q**. We assume the loss of IL-2 is proportional to this new production

**p*q**. Pathway **w** represents the natural loss of IL-2, which is proportional to the level of IL-2 present.

**Parameter Values & Initial Values**

We determined parameter values and initial values through extensive literature research. We estimated any values that were not identifiable in literature. This section details our reasoning for selection of estimated values from the literature, and any necessary calculations we made. In some cases, we use a subscript of $x$ to represent any one of the possible subscripts being considered in a given section.

**Initial Values**

$L_0$: The initial value of liver allograft cells reflects the number of hepatocytes in an average adult human liver weighing between 1.4 and 1.7 Kg - estimated to be 200 billion hepatocytes.[88]

$A_0$: Athanassopoulos et al. serially monitored dendritic cell (DC) counts in venous blood of heart transplant recipients. For recipients in stable condition 38 weeks after transplantation, the last included time point, average dendritic cell concentration was 10.79 ± 1.4 DCs/µL (Athanassopoulos et al., Table 3).[89] Given baseline high antigen load, we estimate that up to 75% of these APCs may be presenting alloantigen. Hence we used 75% of the average DC concentration measured by Athanassopoulos et al. as the initial value of APCs. While immunophenotyping of DC subsets has been completed in liver transplant recipients, relative frequency of cells is reported rather than absolute count. We used DCs to parameterize model APCs counts because DCs generate potent immune responses relative to macrophages or MDSCs.[90]

$T_{R0}$: Treg counts in peripheral blood were measured in 133 kidney transplant recipients 1 year after transplantation by San Segundo et al. The median Treg count was 13.07 cells per uL *(*San Segundo et al., Results).[91] We estimate that up to 10% of circulating Tregs may have alloreactive specificity. Hence we use 10% of the average Treg count measured by San Segundo et al. as the initial value of Tregs. Although lymphocyte subsets have been reported in liver transplant recipients, the ability of such values to inform our parameterization was limited by the short length of time after transplantation at which lymphocyte subsets were measured. Since our model reflects the response to transplantation, after 1 year, initial values must reflect this duration which is critical to the assumptions made in the model reduction phase.

$T_{C0}$: The initial value of cytotoxic T cells was derived from a study by Liyana Abdullah et al. in which peripheral blood lymphocyte subsets of kidney transplant recipients were obtained >12 months after transplantation. In recipients without rejection and with allografts >12 months after transplantation (n=11), cytotoxic T cell counts averaged 997.2 ± 172.71 cells/µL (Liyana Abdullah et al., Table 4).[92] Given that up to 10% of T cells are alloreactive,[93] we use 10% of this measured value as the initial value of cytotoxic T cells. As with Treg initial values, we prioritized matching cell counts with regard to time from transplantation over matching specificity to liver transplantation.

$T_{H0}$: The helper T cell initial value was obtained similar to the method for cytotoxic T cell values as above, from Liyana Abdullah et al. peripheral blood lymphocyte subsets (Liyana Abdullah et al., Table 4).[92]

$I_0$: The starting value of IL-2 is derived from Tefik et al. which reports stable kidney transplant recipients' serum IL-2 concentrations 3 months after transplantation (Tefik et al., Table 3).[94] We note that this time period is significantly shorter than the model setting, and measured from cohort of recipients receiving living, related kidney allografts as limitations of this parameter value. However, serum IL-2 is not routinely monitored. Furthermore, IL-2 abundance is more frequently reported as the percentage of cells staining positive for IL-2, rather than the type of concentration needed for this model.

**Carrying Capacities** $(K_x)$
All carrying capacities were estimated to be six times the initial value.

**Constant Source Rate** $(S_x)$
$S_R$: To calculate the constant source rate of Tregs, we multiplied the number of naive T cells in circulation by the rate of naive T cell differentiation to regulatory T cells. The average adult has approximately $10^{11}$ total naive T cells,[95] about 2% of which are in circulation (in 5.5 L of blood) at any given time.[96] The rate of differentiation of naive T cells to Tregs was obtained from a mathematical model by Thakur et al. which describes the immune dynamics between DCs, regulatory T cells, helper T cells, and cytotoxic T cells involved in in vitro antibody synthesis (Thakur et al., Table 2, $\theta_{Treg}$ mean value of $2.95 \times 10^{-4}$ per day).[97] Importantly, the model by Thakur et al. is reflective of in vitro kinetics and parameters were fit by selecting parameter sets which generated curves reflective of the expected response by each immune cell population in a healthy adult.

**Source Rates Requiring Secondary Interaction**
There are five instances in the defined model in which the source rate, i.e., the number of antigens or cells generated per uL per day, requires a secondary interaction. These sources are defined with equation terms (1) **a** requiring **b**, (2) **m** requiring **g**, (3) **k** requiring **l**, (4) **p** requiring **q**, and (5) **s** requiring **r**. In these Michaelis-Menten style terms, the $\alpha$ parameter in the numerator represents the maximum possible generation rate. The $\beta$ parameter in the denominator represents the population size (cells/μL) at which half of the maximum effect defined by $\alpha$ is observed. For terms **r** in equation 3 and **q** in equation 4, $\alpha$ is missing from the numerator because it is incorporated into the $\gamma$ parameter of the multiplied term.

$\alpha_{AH}$, $\beta_{AH}$: The maximum activation rate of helper T cells by APCs, $\alpha_{AH}$, was taken from the same mathematical model of immune dynamics by Thakur et al. used to estimate the constant source of $T_R$ (Thakur et al., Table 2, $\theta_{Th}$ mean value).[97] We estimate that $\beta_{AH}$, the level of APCs at which half this rate is observed, is 50% of the initial condition, $A_0$.

$\alpha_{HC}$, $\beta_{HC}$: The maximum activation rate of cytotoxic T cells via helper T cell help, $\alpha_{HC}$, was calculated using a study of in vivo dynamics of T cell activation by Ribeiro et al. using deuterated glucose to monitor kinetics.[98] We multiplied the reported mean rate of CD8+ T cell activation among controls, 0.001/day (Ribeiro et al., Table 1, Mean value of *a*), by the number of total CD8+ T cells measured in peripheral blood of transplant recipients more than 1 year after transplantation (approximately 1000 cells/uL; Liyana Abdullah et al.,Table 4).[92,98] We estimate that $\beta_{HC}$, the population of activated CD4+ helper T cells at which half this rate is observed, is 50% of the initial condition, $T_{H0}$.

**Growth Rate Constants** $(\gamma_x)$

Proliferation rate constants may be calculated using the formula: $\gamma_x = \dfrac{ln(2)}{t_D}$, where $t_D$ represents the doubling time. For CD4+ helper T cells, we used a doubling time of 11 hours.[99] For CD8+ cytotoxic T cells, we used a doubling time of 8 hours.[99] Since IL-2 is required for proliferation of conventional T cells, $\gamma_H$ and $\gamma_C$ implicitly incorporate this required dynamic.

**Constants of Proportionality** $(\lambda_x)$
Constants of proportionality are used in interactions in which loss of one model agent is required for the growth of another. $\lambda_L$ represents the average number of APCs primed with alloantigen upon death of one hepatocyte. $\lambda_H$ and $\lambda_C$ are used to indicate the proportional loss (via internalization) of IL-2 required to trigger proliferation of conventional T cell proliferation. $\lambda_R$ represents the proportional loss of IL-2 (also via internalization) necessary to prolong Treg survival.

$\lambda_L$: The number of APCs that become alloreactive through uptake and presentation of graft-derived antigens is not measured and was hence estimated. We estimate that 98% of liver cells undergo cell death through apoptosis, which does not release antigens into circulation, while 2% of cell death occurs via necrosis.[100] We assume that each hepatocyte contains an average of 8.7 × 10⁹ protein molecules[101] which may generate many antigens per protein; we assume 10 potential antigens per protein molecule. However, despite the magnitude of this value, the vast majority of allograft-released proteins share high homology with the donor and thus will not trigger an immune response. Thus, we estimate that only up to 10% of the antigens are alloantigens. We estimate there can be 7 x 10⁶ antigens on a mature APC, but only a fraction (we assume 10%) might be graft antigens.[102] To obtain APCs per µL, we divide by 5.5 x 10⁶ µL. To account for the fact that most antigen presentation happens in the lymphatic system, while our model only includes events that happen in the blood, we divide by a factor of 1000. Altogether, we estimate $\lambda_L$, the number of APCs per µL beginning to present alloreactive antigens per hepatocytes, to be 4.52x10⁻⁹, calculated as follows: $\dfrac{0.02 * 8.7 \times 10^9 * 10 * 0.1}{7 \times 10^6 * 5.5 \times 10^6}$

$\lambda_H$ was estimated based on human T cell culture experiments by Coppola et al. investigating the effect of various cytokines on activated CD4+ helper T cell growth. Based on Coppola et al. Figure 1A, administration of 10 ng/mL of IL-2 (experimental condition 1,0,0) is sufficient to produce the minimum growth rate.[103] We divided this value by the concentration of cells (166,667 cells/mL) and multiply by the relative growth rate 1.05 in (Coppola, Fig. 1A) to roughly approximate the amount of IL-2 necessary per cell to trigger cell division. We acknowledge that the data from this paper and others suggest that the presence of other cytokines may contribute or be sufficient to support T cell proliferation. However, for the purpose of this model, cytokines that serve functions parallel to IL-2 are symbolically represented in our model by IL-2 itself. Please see limitations in the Conclusion section for further discussion of this.

$\lambda_C$ was estimated using data from Cho et al. describing the effects of IL-2 on naive CD8+ T cell activation and growth.[104] To achieve and measure observable effects, Cho et al. cultured CD8+ T cells using 1 μg/mL of IL-2. As we did for the $\lambda_H$ estimation, we divide this concentration of IL-2 by the number of CD8+ T cells in culture to arrive at the quantity of IL-2 required to simulate proliferation per T cell. Since Cho et al. investigated naive CD8+ T cells, which require greater amounts of IL-2 to trigger activation and proliferation than activated CD8+ cytotoxic T cells, we used the upper end of the reported cell culture density range (2 x 10$^5$ cells/well) to account for this, resulting in a smaller constant of proportionality. We assume 200 μL/well.

$\lambda_R$ was estimated based on a study of ex vivo Treg expansion by Hippen et al., which used IL-2 concentration of 300 IU/mL.[105] Using reported specific activity (18.0 x 10$^6$ IU/mg) of the utilized Chiron brand IL-2 to convert to standard units, this concentration equates to 0.0166 ng/μL. Again dividing the IL-2 concentration by the estimated density of cells in culture yields the constant of proportionality. Notably, the density of cells in culture is not reported. However, given that this concentration of IL-2 produced significant expansion of Tregs, we estimate a high density (5.0 x 10$^6$ cells/well) to obtain a constant of proportionality more reflective of the minimum level of IL-2 necessary to prolong Treg lifespan. We assume 200 μL/well. Altogether, we get 6.67x10$^{-7}$, from the following calculation:

$$\frac{(\frac{300 IU}{10^3 \mu L})(\frac{10^6 ng}{18 x 10^6 IU})(200 \mu L)}{(5 \times 10^6 cells)}$$

**Loss Rate Constants** $(\delta_x)$

Loss rate constants may be calculated by taking the inverse of average lifespan, t$_{avg}$: $\delta_x = \frac{1}{t_{avg}}$

or by using the formula: $\delta_x = \frac{ln(2)}{t_{\frac{1}{2}}}$, where t$_{1/2}$ represents half life. The loss rate constant for APCs was calculated by taking the inverse of the average lifespan of circulating dendritic cells, 12 days.[106] Cytotoxic and helper T cell loss rate constants were calculated in the same manner using average lifespans, 41 hours[99] and 3 days respectively.[99] Similarly, natural loss of liver hepatocytes was calculated using average hepatocyte lifespan of 200 days.[107] The IL-2 loss rate constant was calculated using a half life of 6 minutes.[108] Although a second component clearance of 30-120 minutes is reported for IL-2,[108] we assume for the purposes of this model that second component clearance is negligible relative to the magnitude of first component clearance. The Treg loss rate constant was obtained from an in vivo study of Treg kinetics by

Vukmanovic-Stejic et al. which used deuterium labeling to monitor proliferation and disappearance.[109]

**Additional Boost and Inhibition Terms ($\alpha_x$, $\beta_x$)**

$\alpha_{HI}$, $\beta_{HI}$: The maximum activation rate of IL-2 production by helper T cells, $\alpha_{HI}$, was calculated using the same mathematical model of immune dynamics by Thakur et al. used to estimate parameters $S_R$ and $\alpha_{AH}$ (Thakur et al., Table 2, $\alpha_{IL2}$ mean value).[97] We multiplied the parameter value reported by Thakur et al., in units of ng/cell*day, by the initial value of helper T cells, $T_{H0}$ in cells/µL, to approximate the maximum production of IL-2 per day. $\beta_{HI}$ is derived from an in vitro study that demonstrated that paracrine effects of IL-2 secretion, matching the dynamics reflected in this model, are observed at cell concentrations greater than 100 cells/µL.[110]

$\alpha_{IR}$ & $\beta_{IR}$: The maximum effect of IL-2 to prolong Treg lifespan (inhibit natural loss) was estimated from same paper by Hippen et al. that was used to calculate $\lambda_R$. Upon stimulation that included IL-2, nTregs were still functional at 25 days. However, suppressive capacity was not observed when harvested at day 55. Suppressive capacity of cultured nTregs was not reported for timepoints between these days.[105] Hence, we calculate $\alpha_{IR}$ as the fold change between a prolonged average lifespan that we assume is 40 days and the average lifespan of Tregs measured by Vukmanovic-Stejic et al. using deuterated glucose labeling, 15 days.[109] We assume this is a maximal effect, meaning that the loss rate constant for $T_R$ gets multiplied by $(1 - \alpha_{IR})$ to go from 1/15 to 1/40, giving $\alpha_{IR} = 0.625$. We estimate that the threshold concentration of IL-2 needed to achieve half of this effect is 50% of the concentration (so, 50% of 0.0166 ng/µL) used by Hippen et al. to stimulate nTregs.[105] Conversion of reported IU concentration to standard units is as discussed in the $\lambda_R$ parameter description, above.

**Estimated Parameters**

The value of remaining parameters, $\alpha_{CL}$, $\beta_{CL}$, $\alpha_{RA}$, $\beta_{RA}$, $\alpha_{IRA}$, $\beta_{IRA}$, $\alpha_{IH}$, $\beta_{IH}$, $\alpha_{CI}$, $\beta_{CI}$, $\alpha_{IC}$, $\beta_{IC}$ are not measured. To estimate remaining parameters, we chose values we felt were reasonable. Please refer to Table 3 for individual parameter values.

| Table 3: Parameter Descriptions and Nominal Values | | | | | |
|---|---|---|---|---|---|
| # | Parameter | Description | Value | Units | Sources Used |
| 1 | $\delta_L$ | Natural loss rate constant of L | 0.005 | /day | [101] |
| 2 | $\alpha_{CL}$ | Maximum $T_C$ killing effect on L | 10 | - | estimated |
| 3 | $\beta_{CL}$ | Threshold of $T_C$ for half of maximum killing effect on L | 200 | cells/$\mu$L | estimated |
| 4 | $\lambda_L$ | Number of APCs primed by alloantigens, per hepatocyte | 0.00000000452 (4.52x10$^{-9}$) | - | 101 |
| 5 | $\delta_A$ | Natural loss rate constant of APCs | 0.0833 | /day | 107 |
| 6 | $\alpha_{AH}$ | Maximum rate of activation of T$_H$ by APCs | 0.0000261 | cells/$\mu$L*day | 97 |
| 7 | $\beta_{AH}$ | Threshold of APCs for half of maximum T$_H$ activation effect | 4 | cells/$\mu$L | 97 |
| 8 | $\alpha_{RA}$ | Maximum rate of Treg suppression of APC presentation | 0.4 | - | estimated |
| 9 | $\beta_{RA}$ | Threshold for half of maximum rate of Treg suppression of APC presentation | 20 | cells/$\mu$L | estimated |
| 10 | $\alpha_{IRA}$ | Maximum boost of IL-2 on Treg suppression of APC presentation | 2 | - | estimated |
| 11 | $\beta_{IRA}$ | Threshold of IL-2 for half of maximum suppression of APC presentation | 0.356 | ng/$\mu$L | estimated |
| 12 | $\gamma_H$ | Proliferation rate constant for logistic growth of T$_H$ | 1.51 | /day | 99 |
| 13 | $K_H$ | Carrying capacity for T$_H$ | 422 | cells/$\mu$L | 92 |
| 14 | $\alpha_{IH}$ | Maximum effect of IL-2 on T$_H$ proliferation | 2 | - | estimated |
| 15 | $\beta_{IH}$ | Threshold of IL-2 for half of maximum proliferation effect on T$_H$ | 0.178 | ng/$\mu$L | estimated |
| 16 | $\delta_H$ | Natural loss rate constant of T$_H$ | 0.333 | /day | 99 |
| 17 | $\alpha_{HC}$ | Maximum activation rate of T$_C$ by T$_H$ | 1 | cells/$\mu$L*day | 98 |
| 18 | $\beta_{HC}$ | Threshold for half maximum activation of T$_C$ by T$_H$ | 35 | cells/$\mu$L | 98 |
| 19 | $\gamma_C$ | Proliferation rate constant for | 2.08 | /day | 99 |
| | | logistic growth of $T_C$ | | | |
| 20 | $K_C$ | Carrying capacity for $T_C$ | 598 | cells/$\mu$L | 92 |
| 21 | $\alpha_{IC}$ | Maximum effect of IL-2 on $T_C$ proliferation | 2 | - | estimated |
| 22 | $\beta_{IC}$ | Threshold for half of maximum effect of IL-2 on $T_C$ proliferation | 0.178 | ng/$\mu$L | estimated |
| 23 | $\delta_C$ | Natural loss rate constant of $T_C$ | 0.585 | /day | 99 |
| 24 | $s_R$ | Constant source rate of Tregs | 0.107 | cells/$\mu$L*day | 95–97 |
| 25 | $\delta_R$ | Natural loss rate constant of Tregs | 0.0658 | /day | 109 |
| 26 | $\alpha_{IR}$ | Maximum lengthening effect IL-2 has on Treg lifespan | 0.625 | - | 105,109 |
| 27 | $\beta_{IR}$ | Threshold of IL-2 for half of maximum change in Treg lifespan | 0.00833 | ng/$\mu$L | 105 |
| 28 | $\alpha_{CI}$ | Maximum rate of IL-2 production by $T_C$ | 0.36 | ng/$\mu$L*day | estimated |
| 29 | $\beta_{CI}$ | Threshold for half of maximum of IL-2 production by $T_C$ | 352 | cells/$\mu$L | estimated |
| 30 | $\alpha_{HI}$ | Maximum rate of IL-2 production by $T_H$ | 70.7 | ng/$\mu$L*day | 97 |
| 31 | $\beta_{HI}$ | Threshold for half of maximum of IL-2 production by $T_H$ | 99.7 | cells/$\mu$L | 110 |
| 32 | $\lambda_C$ | Rate and proportionality constant for the number of IL-2 molecules required for production of 1 $T_C$ cell | 0.001 | ng/cell | 104 |
| 33 | $\lambda_H$ | Rate and proportionality constant for the number of IL-2 molecules required for production of 1 $T_H$ cell | 0.0000063 | ng/cell | 103 |
| 34 | $\lambda_R$ | Proportionality constant of IL-2 for lifespan fold change of Treg | 0.000000667 | ng/cell | 105 |

| 35 | $\delta_I$ | Natural loss rate constant of IL-2 | 166 | /day | 108 |

## Results

**Simulations**

After parameterizing the model, we put the model and parameter values into MATLAB v. R2024a and plotted solutions to the system. All code is available in the Supplementary Materials. In our simulations, the beginning time of the simulation is labeled time zero, which is approximately one year after transplant. Our simulation runs over a time period of thirty days starting from time zero, with the assumption that the process of rejecting the allograft organ is initiated at time zero.

**Liver:** We assume that hepatocytes begin in equilibrium at their carrying capacity, and as time starts, cytotoxic T cells begin to induce apoptosis through the granzyme/perforin pathway.

**APCs:** We assume the initial value of APCs is at a low baseline level. As the liver releases alloantigens triggering the recipient immune response, cytotoxic-mediated cell damage augments this process. Increased antigen load with damage to the allograft increases the number of APCs that encounter allograft antigen, phagocytose, and present alloantigens.

**Helper T cells:** We assume the initial value of activated alloreactive helper T cells is low. As the population of alloantigen-presenting APCs grows, more alloreactive helper T cells are activated. Helper T cells also grow due to IL-2 stimulating their proliferation, which results in a positive feedback loop as helper T cells produce IL-2.

**Cytotoxic T cells:** We assume the initial value of activated cytotoxic T cells is low, and the population grows due to IL-2 stimulating their proliferation. In addition, activated helper T cells continue to activate more alloreactive cytotoxic T cells.

**Regulatory T cells:** We assume the initial value of Tregs is low, and they slowly grow over time due to the slow constant source we have assumed. Also, since IL-2 levels are high, T reg cells' lifespans are lengthened, so fewer will die over time. These two mechanisms combined ultimately lead to slow growth of the T reg population.

**IL-2:** As helper T cells and cytotoxic T cells rapidly grow, so does IL-2. We believe the initial spike seen in the growth is due to the small initial value assumed, and then there is slower, but still rapid, growth of IL-2 likely due to the production by the helper T cells and cytotoxic T cells, until it reaches an equilibrium value.

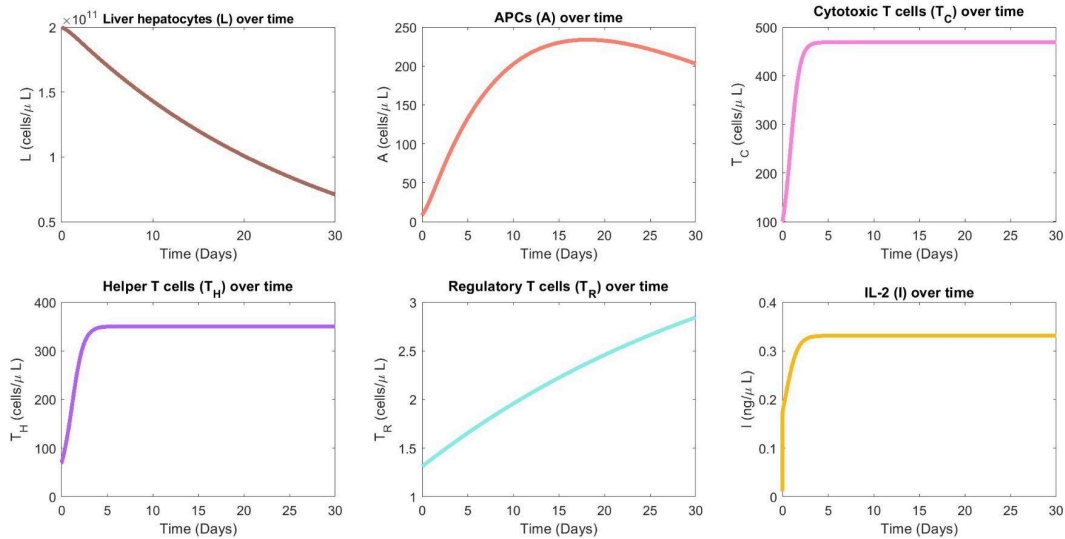

**Figure 2: Simulations of graft-immune dynamics during acute rejection.** Simulations were run over a thirty-day period of acute rejection, using nominal parameter values (Table 3).

**Sensitivity Analysis**

We used MATLAB to conduct a global sensitivity analysis using the Sobol' method. For each parameter listed in the table, we constructed ranges of feasible values by sampling between 50% and 150% of the nominal values. We used uniform distributions for each parameter, and used the Sobol' sequence sampling scheme. A plot of the first-order and total Sobol' indices is shown in **Fig. 3**, which omits the other 25 parameters with smaller total Sobol' indices. We looked for the parameters with the largest total Sobol' indices. We could have selected the top 4 or the top 7, as each of these parameter groups have indices an order of magnitude larger than the remaining indices. We selected 7 to try to capture as much of the full-model behavior as possible. The top 7 most-influential parameters are the following: $\delta_C, \delta_L, \alpha_{CL}, \beta_{CL}, \delta_I, \alpha_{IC}, \gamma_C$. To see whether using just these 7 is sufficient to capture most of the full-model behavior, we compared the probability density functions (PDFs) of QOI values when samples are generated while varying all 35 parameters versus when just these 7 most-influential parameters are varied (**Fig. 4a**). The concordance of the PDFs indicates that if we only vary the 7 most-influential parameters, we obtain almost all of the variability present when all 35 parameters are varied. In contrast, **Fig. 4b** compares the PDF with all 35 parameters varied with the PDF when only the remaining 28 least-influential parameters are varied. It is apparent that varying the 28 remaining parameters captures only a very small part of the QOI-value range, and much less than varying the 7 most-influential parameters does.

We note that the 7 most-influential parameters represent rates involved in the dynamics of the cytotoxic T lymphocytes, the healthy liver hepatocytes, and IL-2. The most-influential parameters are those that have the biggest impact on the QOI [which is L(30), the number of healthy liver hepatocytes at day 30] when the parameter values are changed. Thus they can be used to narrow down the search for potential novel therapies, or to inform the selection of therapies for optimal combinations. The 7 most-influential parameters are represented graphically using a heat map shown in **Fig. 5**.

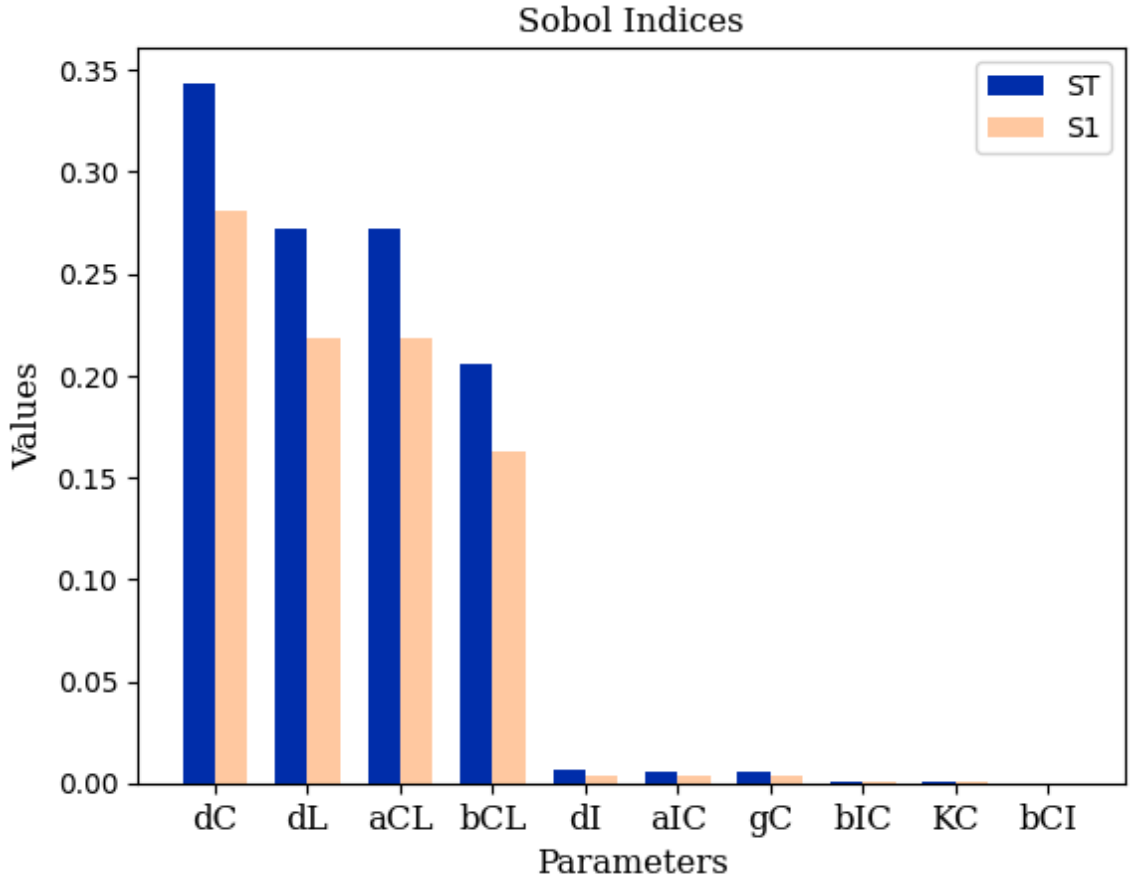

**Figure 3: Global sensitivity analysis bar chart, ordered by total Sobol' index values.** We used Python 3.10, accessed from Google Colab, to plot the Sobol' indices obtained from our global sensitivity analysis in MATLAB. The Sobol' indices were obtained using 100,000 base samples, and after calculating the first-order index, we ended up with 3,700,000 samples. In descending order of influence, the most-influential parameters (on the horizontal axis in English letters) correspond to $\delta_C$, $\delta_L$, $\alpha_{CL}$, $\beta_{CL}$, $\delta_I, \alpha_{IC}, \gamma_C$. We consider the other twenty-eight parameters to be non-influential.

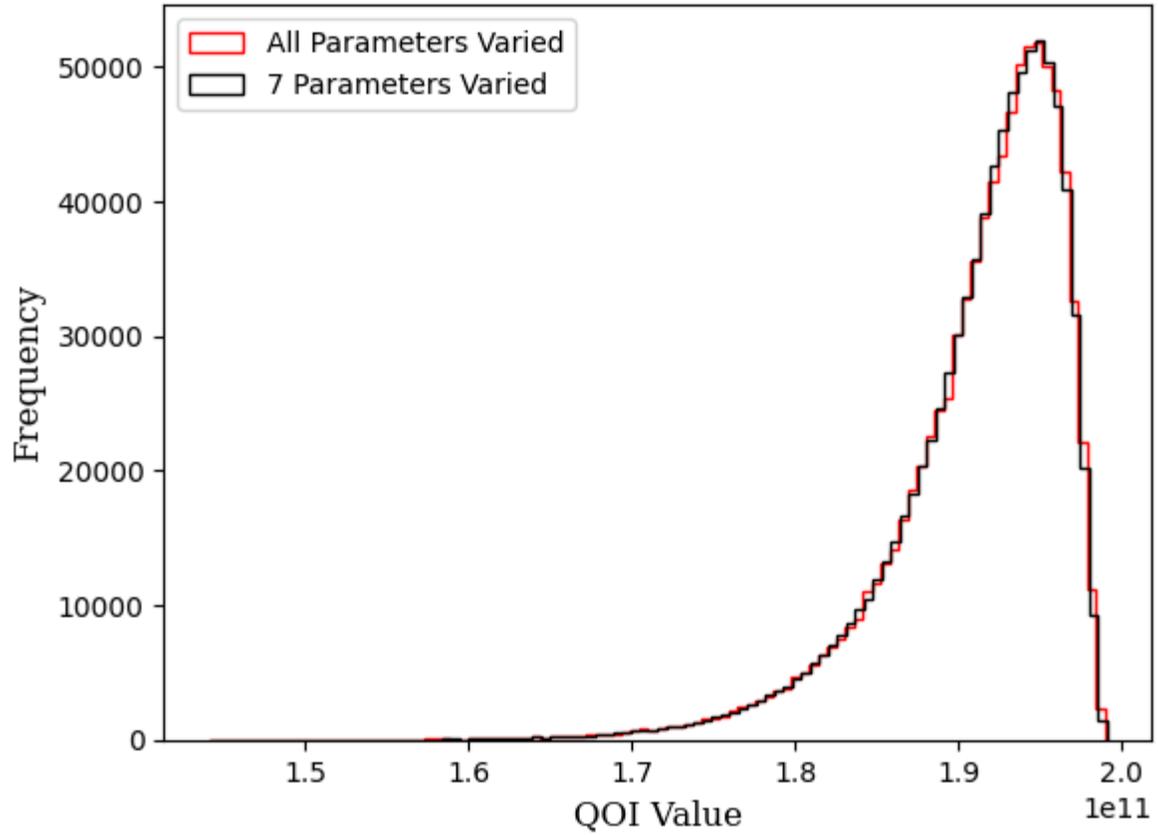

**4a**

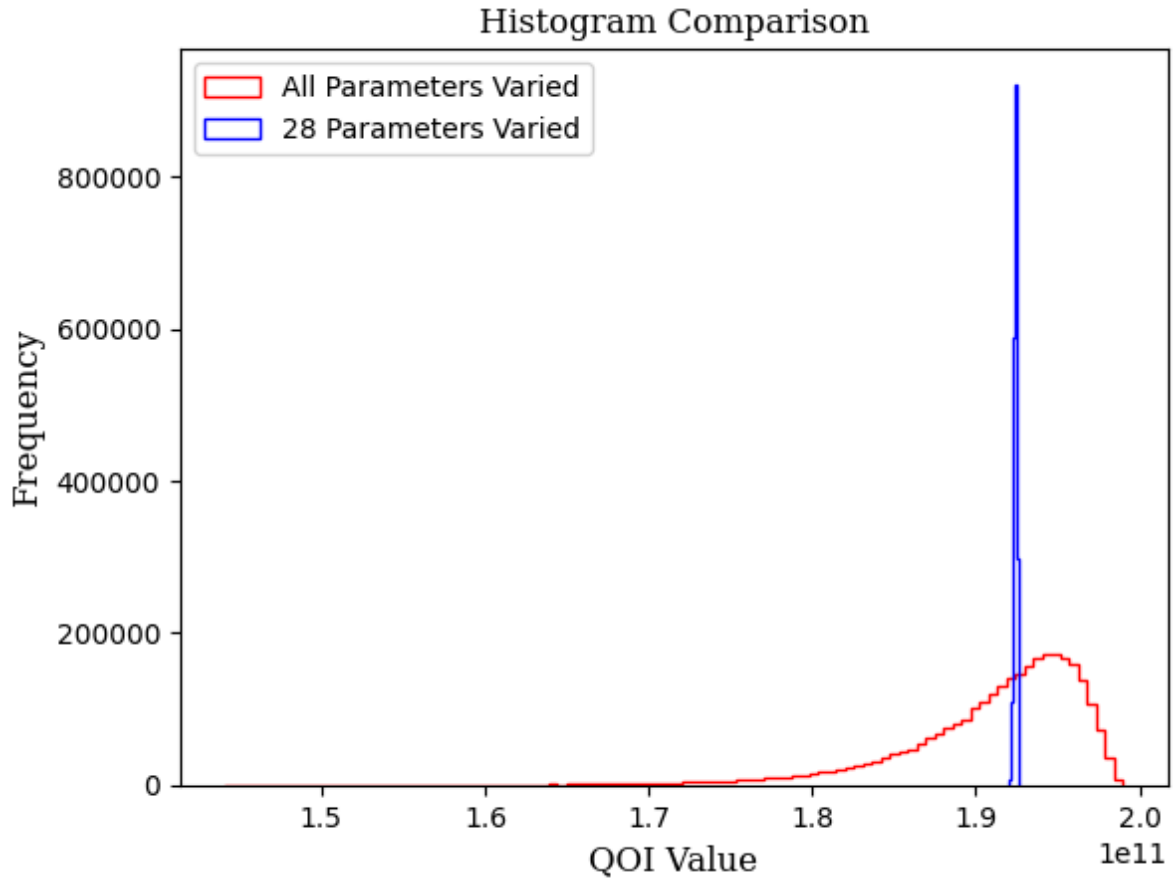

**4b**

**Figure 4: Variability in QOI values for different parameter distributions. (a)** Histograms of the QOI values obtained while sampling parameter space by varying all parameters (red outline), and then just with the 7 most-influential parameters varying (black outline). The two histograms were generated using the number of samples that were needed for Sobol' indices sampling of the 7 most-influential parameters. Calculating the Sobol' indices with only 7 parameters varying yields ((7+2)*100,000) = 900,000 samples, so we compared this to a random subset of 900,000 from the 3,700,000 samples obtained from letting all 35 parameters vary. The histogram from varying the 7 most-influential parameters is very close to the histogram with all 35 parameters varying, indicating that the 7 parameters capture almost all of the variability seen from the full 35 parameters. **(b)** Histograms of the QOI values obtained from varying all parameters (red outline) and varying only the twenty-eight least-influential parameters (red outline). Varying 28 parameters, with 100,000 base samples used for the Sobol' index calculations, yields 3,000,000 samples, so we selected a random subset of 3,000,000 samples of the 3,700,000 samples obtained by letting all parameters vary. The twenty-eight least-influential parameters produce almost none of the variability in the QOI value that is obtained when varying all 35. We used Python 3.10 accessed through Google Colab to generate the plots in this figure.

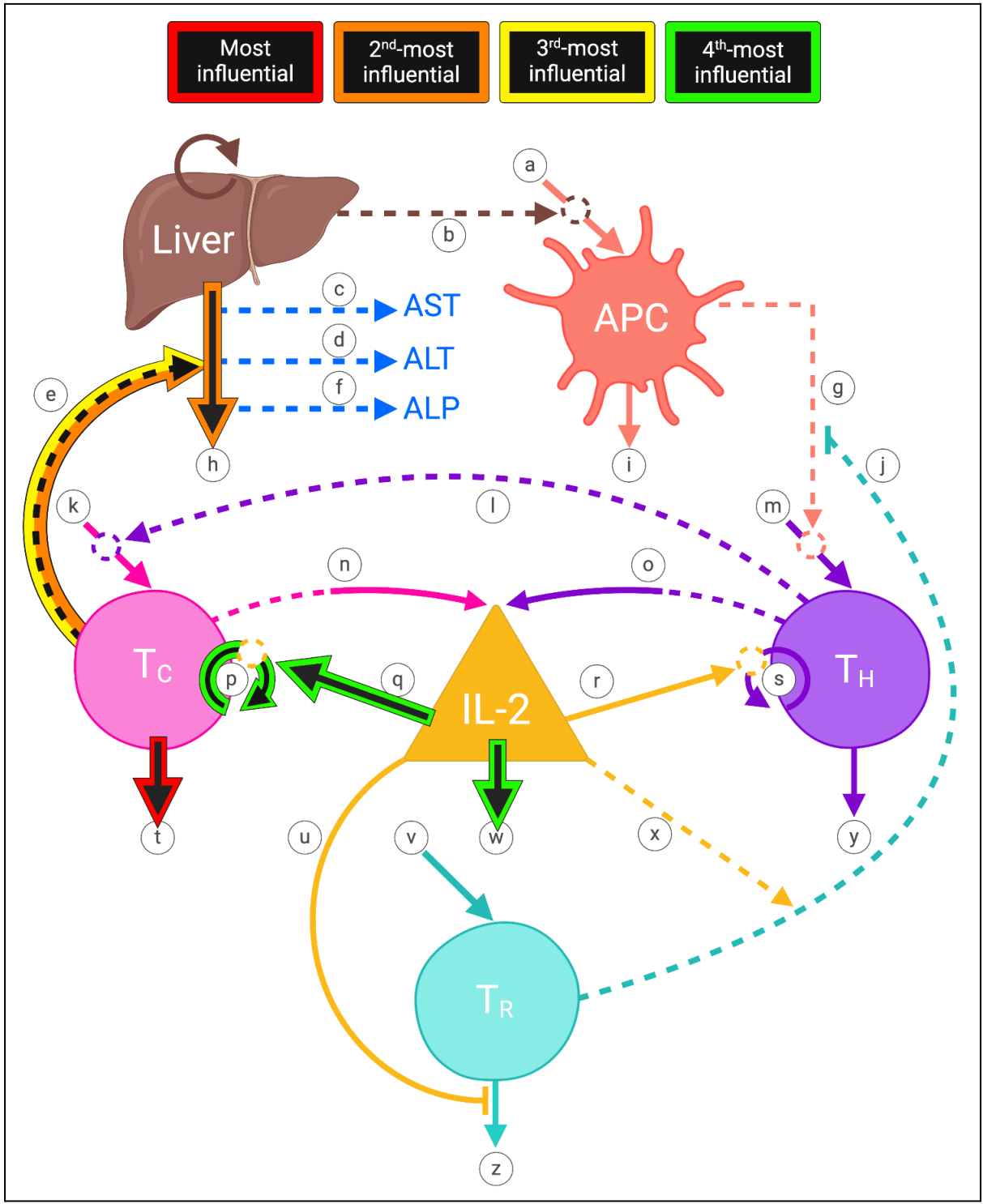

**Figure 5:** Heat map of sensitivity analysis results. Pathways with parameters that are highly influential on the QOI (level of liver hepatocytes at day 30) are highlighted in colors reflecting the relative level of influence. Pathways that are not highlighted were found to be much less influential.

## Discussion

Patients with end-stage liver disease can gain additional years of healthy lifespan by undergoing liver transplantation. However, significant morbidity and mortality can still occur from over-suppression or under-suppression of the immune system, from the immunosuppressive medications patients take to prevent graft rejection. A better understanding of organ graft and immune system dynamics would provide insights into ways to control immunosuppression, leading to lower morbidity and mortality in patients. In this work, we developed a mechanistic model of liver graft transplant and immune cell dynamics, to study the dynamics of rejection. We began with an extensive literature search of mechanisms related to liver transplant rejection, or organ transplant rejection more generally. This resulted in a large number of immune cell populations, cytokines, and interactions in our initial model draft. We reduced that model by including only the most critical cells and cytokines, resulting in the model presented in this work. This reduction allowed us to use data from the literature to estimate all but a few parameters in the model.

In our current model of immune dynamics and organ transplant rejection, we have made various assumptions. One limitation is that all of the cell types, cytokines, and immune interactions we include in our model represent dynamics in blood. A three-compartment model representing immune dynamics within the blood, lymph, and the allograft itself, albeit considerably more complex, may more closely match the reality of a complex process such as allograft rejection. Nonetheless, our model here represents a substantial and novel first step in that direction, as well as a robust representation of the key features of liver allograft rejection in the blood. Our model is also limited by the availability of parameter values, many of which have not been rigorously measured by the scientific and medical community and thus had to be estimated. Conversely, this highlights some important knowledge gaps in the field of liver allograft rejection and the need for rigorous basic or translational studies to fill them.

Additionally, we omitted a number of potential additional model features. We excluded innate immune cells from our model. Since our setting is one year after transplantation, we assume the level of damage-associated molecular patterns (DAMPs), which peaks in the early postoperative setting due to allograft ischemia and surgical trauma, reaches a baseline level that is negligible relative to stimulation of the adaptive immune system. We also excluded B cells, as T cell-mediated rejection (TCMR) is much more common than antibody-mediated rejection (ABMR) in the liver transplant setting.[111] Furthermore, the included T cell populations represent broad subsets rather than particular lineages. For example, helper T cells are represented as a general single group rather than separated as Th1, Th2, Th17, etc. This is due to uncertainty in their individual roles. Similarly, regulatory T cells are represented as a single group, rather than separated into iTregs and nTregs. Finally, the only cytokine explicitly included in this model is IL-2. Limited cell and cytokine inclusions mean that other factors involved in the immune dynamics of allograft transplant will be represented implicitly in any fitted parameters for those factors included in the model. Once we fit our model to data, it will be important to acknowledge the role of additional factors not included in our model in the fitted parameter estimates.

We simulated our model with the nominal parameter values, to obtain the general behavior shown in **Fig. 2**. Using Sobol' global sensitivity analysis, we found that $\delta_C$, $\delta_L$, $\alpha_{CL}$, $\beta_{CL}$, $\delta_I$, $\alpha_{IC}$, $\gamma_C$ are the most-influential parameters on the number of viable hepatocytes (**Fig. 3**). These parameters are involved in the dynamics of the cytotoxic T lymphocytes, the healthy liver hepatocytes, and IL-2. These influential pathways are highlighted in the heat map in **Fig. 5**, and are pathways that are predicted to be able to make the largest difference in our QOI (hepatocyte levels on day 30 after initiation of allograft rejection). The identification of these key parameters has significant implications for the use of diagnostic tests to monitor patients, and therapeutic strategies to prevent or treat liver transplantation rejection. The influential parameters highlight critical aspects of the immune response and cellular dynamics in the graft, providing actionable targets for intervention. The most-influential parameters may represent novel potential therapeutic targets, or their signaling pathways/dynamics could be explored further to identify such targets. The most-influential parameters can be used to decide which therapies to use or combine, to best control the rejection.

Currently-used immunosuppressive medications such as calcineurin inhibitors (e.g., tacrolimus and cyclosporine) are cornerstones in post-transplant care.[112,113] These inhibitors block the activation of T cells by inhibiting the calcineurin pathway, effectively reducing T cell proliferation and cytotoxic activity.[114] This aligns with the identified importance of the $T_C$ proliferation and the $T_C$ killing effect on hepatocytes. mTOR inhibitors (e.g., sirolimus and everolimus) inhibit a pathway crucial for T cell proliferation and function.[112,114] These agents help modulate the immune response by directly influencing T cell dynamics, which is critical given the model's sensitivity to the rate of $T_C$ proliferation.[114] Corticosteroids such as prednisone and methylprednisolone serve as broad-spectrum immunosuppressants, reducing the overall immune cell activity and inflammation.[114] This non-specific suppression can counterbalance elevated T cell activity and other immune responses that compromise graft survival. Monoclonal antibodies such as basiliximab and daclizumab target the IL-2 receptor on T cells, thereby inhibiting IL-2 mediated T cell activation and proliferation.[112,114] These therapeutics directly address the parameter related to the upregulatory effect of IL-2 on $T_C$ proliferation, which is pivotal in mitigating acute immune responses against grafted tissue. Potential future developments might include specific anti-IL-2 therapies that could finely tune IL-2 levels and their immune activation timing, thereby offering precise control over T cell expansion and function.

The development of diagnostic tests focusing on biomarkers such as IL-2 levels and T cell activity could enable early detection of rejection. The high sensitivity of the model to IL-2 levels and T cell activity suggests that measuring these biomarkers could provide early indications of an impending rejection episode. Techniques such as enzyme-linked immunosorbent assays (ELISAs) and flow cytometry could quantify IL-2 concentrations and T cell subsets, respectively, offering real-time insights into the immunological status of the graft. These would go hand-in-hand with standard-of-care liver function tests that routinely monitor hepatocyte viability and turnover and are currently the only alert clinicians get for graft injury.

These highly-influential parameters can also be used further to personalize the model to data from specific patients, creating medical digital twins.[115] These digital twins can then be studied further in silico, testing various interventions, and even conducting large numbers of in silico clinical trials to study patient-to-patient variability. It is also possible to use optimal control to calculate the best intervention strategies for a digital twin,[21,116] or robust optimal control to calculate best interventions when uncertainty is significant, for either one or a collection of digital twins.[117] Once simulations and calculations are complete, predictions can be validated by testing in clinical trials, with arms for the predicted best regimens.

Our model is of liver allograft transplant rejection, but many of our model features could be used to model transplant rejection of other organs, or possibly even other immune-mediated diseases. Assuming the cells we include in the immune dynamics are relevant for other organs or tissues, our model could be tailored to other settings by making any necessary updates to the parameter values. Changes in relevant cell types can also be made, by adding or removing equations as needed. Then, simulations, sensitivity analysis, and in silico testing of interventions could all be performed for additional organs or tissues. We have included all of our code, with extensive commenting, in the Supplementary Materials. Our goal in this work is to provide the community with a foundation of accessible tools that can be used to improve transplant patient outcomes.

## Acknowledgements

We would like to acknowledge and thank Dr. Jaimit Parikh for providing the code we used for our Sobol' sensitivity analysis, and for his time spent helping us debug our implementation of it.

**Conflicts of Interest:** The authors have no relevant conflicts of interest for this work.

**Code**: The code used to produce all figures can be accessed at:
https://github.com/KyleAdams26/LTCode